\documentclass[aps,showpacs,pra,twocolumn,superscriptaddress]{revtex4}%
\usepackage{multirow}
\usepackage{color}
\usepackage{amsfonts}
\usepackage{amsmath}
\usepackage{mathrsfs}
\usepackage{amssymb}
\usepackage{wasysym}
\usepackage{graphicx}
\begin{document}
%
\title{Tuning the Quantum Phase Transition of Bosons in Optical Lattices\\ via Periodic Modulation of s-Wave Scattering Length}
\author{Tao Wang}
\email{tauwaang@gmail.com} 
\affiliation{Physics Department and Research Center OPTIMAS,
University of Kaiserslautern, 67663 Kaiserslautern, Germany}
\affiliation{Hanse-Wissenschaftskolleg, Lehmkuhlenbusch 4, 27733
Delmenhorst, Germany}
\affiliation{Department of Physics,
Harbin Institute of Technology, Harbin 150001, China}
\author{Xue-Feng Zhang}
\affiliation{Physics Department and Research Center OPTIMAS,
University of Kaiserslautern, 67663 Kaiserslautern, Germany}
\author{Francisco Ednilson Alves dos Santos}
\affiliation{Instituto de F\'{\i}sica de S\~{a}o Carlos, USP, Caixa Postal 369, 13560-970 S\~{a}o Carlos, S\~{a}o Paulo, Brazil}
\author{Sebastian Eggert}
\affiliation{Physics Department and Research Center OPTIMAS,
University of Kaiserslautern, 67663 Kaiserslautern, Germany}
\author{Axel Pelster}
\email{axel.pelster@physik.uni-kl.de} \affiliation{Physics
Department and Research Center OPTIMAS, University of
Kaiserslautern, 67663 Kaiserslautern, Germany}
\affiliation{Hanse-Wissenschaftskolleg, Lehmkuhlenbusch 4, 27733
Delmenhorst, Germany}
\begin{abstract}
We consider interacting bosons in a 2D square and a 3D cubic optical lattice with a periodic
modulation of the s-wave scattering length. At first we map the underlying periodically driven Bose-Hubbard model for large enough driving frequencies
approximately to an effective time-independent Hamiltonian with a conditional hopping. Combining different analytical approaches with quantum Monte Carlo
simulations then reveals that the superfluid-Mott insulator quantum phase transition still exists despite the periodic driving and
that the location of the quantum phase boundary turns out to depend quite sensitively on the driving amplitude. A more detailed quantitative analysis
shows even that the effect of driving can be described  within the usual Bose-Hubbard model provided that the hopping is rescaled appropriately with the driving amplitude.
This finding indicates that the Bose-Hubbard model with a periodically driven s-wave scattering length
and the usual Bose-Hubbard model belong to the same universality class from the
point of view of critical phenomena.
\end{abstract}
\maketitle
\section{Introduction}
Systems of ultracold bosonic gases in optical lattices represent nowadays
a popular research topic
\cite{Fisher:PRB89,Jaksch:PRL98,Greiner:Nat02,Lewenstein:AdP07,Bloch:RMP08,Sanpera,Inguscio},
as they establish a versatile bridge between the field of
ultracold quantum matter and solid-state systems
\cite{Bloch:NaP12}. 
In particular, they can be experimentally
controlled with a yet unprecedented level of precision
\cite{Bloch:NaP05}. With this it is now possible to achieve strong
correlations and, due to the absence of impurities, they are
viewed as idealized condensed matter systems, which allow for a
clear theoretical analysis \cite{Zoller:AP(NY)05} and which are
even predestined as universal quantum simulators \cite{feynman}.

Recently a new degree of freedom to tune the properties of quantum matter has emerged which is based on a real-time modulation of some lattice parameter.
For instance, it was predicted in Ref.~\cite{oldenburg} and later on confirmed experimentally in Refs.~\cite{ex1,ex2}
that the Bose-Hubbard model for a periodically shaken optical lattice
can be approximately reduced for a sufficiently large frequency to an effective time-independent Bose-Hubbard model with a renormalized hopping parameter. This technique
was recently applied to induce dynamically the Mott-insulator (MI) to superfluid (SF) transition \cite{pisa}. Other proposed
applications concern the quantum simulation of frustrated classical magnetism 
as well as the generation of abelian and non-abelian gauge fields \cite{andre1,andre2,andre3,andre4}. Recently, this line of research culminated in the experimental realization of the 
Hofstadter or Harper Hamiltonian with ultracold atoms in optical lattices \cite{bloch,ketterle}.

Another method to periodically drive an ultracold quantum gas system relies on a periodic modulation of the s-wave scattering length
\cite{bagnato1,bagnato2}, which can be experimentally achieved, for instance, in the vicinity of a broad Feshbach resonance \cite{hulet}.
For a harmonically trapped Bose-Einstein condensate this induces various phenomena of nonlinear dynamics as, for instance, mode coupling, higher harmonics generation, and
significant shifts in the frequencies of collective modes (see, for instance, Ref.~\cite{hamid} and the references therein). Therefore, 
a periodic modulation of the interaction represents an important new tool
for building more versatile quantum simulators.

In this work we investigate the effect of a 
periodic modulation of the s-wave scattering length for bosons in an optical lattice.
Using the Floquet formalism we show that the underlying driven Bose-Hubbard model
can be understood for large enough driving frequencies in terms of an effective 
conditional hopping in the sense that its value depends on the respective 
particle numbers of the involved neighboring sites \cite{luis}. Such 
conditional hopping amplitudes represent an interesting class of models, which 
have an intricate history in condensed matter physics.
Correlated or conditional hopping already appeared, for instance,
in the very first paper, where the Hubbard model was proposed, due
to matrix elements of the Coulomb interaction between nearest
neighbor Wannier wave functions \cite{condhopp1}. However, such
occupation-dependent hopping terms are usually smaller than the
typical hopping and on-site interaction, so they can be neglected.
Later on, such terms were reconsidered as an alternative scheme
for high-$T_c$ superconductivity \cite{condhopp2,condhopp3}. For
particular parameter values the Hamiltonian for a Hubbard chain
turns out to be integrable \cite{condhopp4} and displays
fractional statistics \cite{condhopp5}. The occupation-number
sensitivity of tunnelling was even implemented experimentally by
employing the high-resolution quantum gas microscope technique
\cite{condhopp6}. Whereas all these works deal with real-valued
conditional hopping terms, a density-dependent Peirls phase was
recently proposed in Ref.~\cite{anyons} in the context of
realizing anyons in one-dimensional lattices.   One goal of this work is to 
develop analytical tools in order to study the the quantum phase diagram
of conditional hopping models in quantitative detail.

In the following we focus on the case of a
2D square and a 3D cubic lattice, where the conditional hopping allows to tune 
the superfluid-Mott insulator quantum phase transition.
Consequences for the one-dimensional lattice problem have already been discussed in 
detail in Ref.~\cite{luis}, where
a large enough driving induces pair superfluidity.
In Section \ref{section2} we review in detail how the periodically driven 
Bose-Hubbard model reduces approximately to an effective time-independent model
with conditional hopping. Afterwards, we combine in Section \ref{section3} different analytical approaches with quantum Monte-Carlo (QMC) simulations in order to determine 
how the MI-SF quantum phase boundary depends on the driving amplitude.  A more quantitative analysis in Section \ref{section4}
shows that the effect of driving can even be described  within the usual Bose-Hubbard model provided that the hopping is rescaled appropriately with the driving amplitude.
This finding indicates that the Bose-Hubbard model with a periodically driven s-wave scattering length
and the usual Bose-Hubbard model belong to the same universality class from the
point of view of critical phenomena.
\section{Model}\label{section2}
We start with
the derivation that a Bose-Hubbard Hamiltonian with a periodic modulation of the s-wave scattering length can approximately be mapped for large driving frequencies
with the help of Floquet theory to an
effective time-independent Hamiltonian with a conditional hopping.
\subsection{Time-Dependent Hamiltonian}
In the following we study a system of spinless bosons in a
homogeneous lattice of arbitrary dimension $D$ with a periodically modulated s-wave scattering length
\cite{bagnato1,bagnato2}, which can be experimentally achieved, for instance,
in the vicinity of a broad Feshbach resonance \cite{hulet}. This periodically driven quantum many-body
system is described by the time-dependent Hamiltonian 
\begin{eqnarray}
\hat{H}(t)&=&\sum_{i} \left\{ \frac{1}{2}  \Big[ U + A  \cos\left(\omega t\right) \Big]  \left(\hat{n}_{i}^{2}-\hat{n}_{i}\right) -\mu \hat{n}_{i} \right\}
\nonumber \\& &
-\sum_{ij}t_{ij}\hat{a}^{\dag}_{i}\hat{a}_{j} 
\,.
\label{H}
\end{eqnarray}
Here $\hat{a}^{\dag}_{i}$, $\hat{a}_{j}$ denote the
annihilation, creation operators fulfilling the standard bosonic
commutator relations and $\hat{n}_{i}=\hat{a}^{\dag}_{i}\hat{a}_{i}$
represents the particle number operator.
Furthermore, $t_{ij}$ stand for the
respective hopping matrix elements between the sites $i$ and $j$, which are usually
non-zero only for nearest neighboring sites
$i$ and $j$ with $t_{ij}=t$. The local time-independent part depends on
the repulsive on-site energy $U$ as well as on the chemical potential $\mu$ due to the grand-canonical description.
Furthermore, the periodic modulation of the s-wave scattering length is described by the amplitude $A$ and the frequency $\omega$.
Thus, the external driving leads to a quadratic dependence on the
particle number operator, whereas a shaken optical lattice only involves
a corresponding linear dependence \cite{martin1}.
\subsection{Floquet Basis}
For the sake of generality we observe that the Hamiltonian  (\ref{H}) is of the form
\begin{eqnarray}
\hat{H}(t)=\sum_{i}\Big[ f \left( \hat{n}_{i} \right) +Ag \left(\hat{n}_{i}\right)\cos\left(\omega t\right)\Big]
-\sum_{ij}t_{ij}\hat{a}^{\dag}_{i}\hat{a}_{j} 
\,,
\label{HH}
\end{eqnarray}
where the local time-independent part reads
\begin{equation}
\label{f}
f (\hat{n}_{i})=\frac{U}{2} \left(\hat{n}_{i}^{2}-\hat{n}_{i}\right)-\mu \hat{n}_{i}\, ,
\end{equation}
and the local time-dependent part is described by the operator
\begin{equation}
\label{g}
g_{i}(\hat{n}_{i})=\frac{1}{2}\left(\hat{n}^{2}_{i}-\hat{n}_{i}\right) \, .
\end{equation}
In the following we perform a detailed analysis of the general Hamiltonian (\ref{HH}) with arbitrary operators $f (\hat{n}_{i})$ and $g (\hat{n}_{i})$.
In view of choosing a suitable basis, we start with
collecting all local terms in the unperturbed Hamiltonian
\begin{eqnarray}
\hat{H}_{0}(t)=\sum_{i}\Big[ f(\hat{n}_{i})
+Ag_{i}(\hat{n}_{i})\cos\left(\omega t\right)\Big] \,,\hspace*{1mm}
\label{H0}
\end{eqnarray}
which fulfills the periodicity condition
\begin{equation}
\label{PH}
\hat{H}_{0}(t)=\hat{H}_{0}(t+T)
\end{equation}
with period $T=2\pi/\omega$.
Therefore, we can use the Floquet theory \cite{sambe,martin1} which states that the Schr\"odinger equation
\begin{equation}
i \hbar \frac{\partial}{\partial t} | \psi (t) \rangle = \hat{H}_0(t)| \psi (t) \rangle
\end{equation}
has Floquet solutions
\begin{equation}
| \psi_{\alpha} (t) \rangle = | \alpha (t) \rangle e^{- i \varepsilon(\alpha) t / \hbar}
\end{equation}
with some quantum number $\alpha$,
where the Floquet functions $| \alpha (t) \rangle$ have the same periodicity (\ref{PH}) as the unperturbed Hamiltonian $\hat{H}_0(t)$,
i.e.
\begin{equation}
\label{P}
| \alpha (t) \rangle = | \alpha (t+T) \rangle \,.
\end{equation}
These Floquet functions $| \alpha (t) \rangle$ and the corresponding quasienergies $\varepsilon(\alpha)$ thus fulfill the eigenvalue problem
\begin{equation}
\label{EV}
\hat{\mathcal{H}}_{0} (t) | \alpha (t) \rangle = \varepsilon(\alpha) \, | \alpha (t) \rangle
\end{equation}
with the corresponding Floquet Hamiltonian
\begin{equation}
\hat{\mathcal{H}}_{0}(t)=\hat{H}_{0}(t) -i\hbar\frac{\partial}{\partial t} \,.
\end{equation}
In order to solve the eigenvalue problem (\ref{EV}) of the unperturbed Hamiltonian (\ref{H0}), one introduces an extended Hilbert space \cite{martin1},
in which the time $t$ is explicitly considered
as a separate coordinate. Two $T$-periodic
functions $|u_{1}(t)\rangle$ and $ |u_{2}(t)\rangle$, which have the scalar product  $\langle u_{1}(t)|u_{2}(t)\rangle$ in the usual Hilbert space,
then have a modified  scalar product in the extended space which reads
\begin{equation}
\label{SC}
\langle\langle u_{1}(t)|u_{2}(t)\rangle\rangle = \frac{1}{T}\int_0^T dt \,\langle u_{1}(t)|u_{2}(t)\rangle \,.
\end{equation}
Now in the extended Hilbert space, the eigenvalue problem (\ref{EV}) can be solved exactly within the occupation number representation.
This yields the Floquet functions
\begin{eqnarray}
\label{SOL}
|\{n_{i}\} , m (t) \rangle\rangle = e^{i m \omega t}  \prod_{i} \left[ e^{- \frac{i A g(n_i)}{\hbar \omega} \sin (\omega t)}  |n_{i}\rangle_i \right] \, ,
\end{eqnarray}
where $|n_{i}\rangle_i$ represents the occupation number basis at site $i$, 
and the Floquet eigenvalues read
\begin{eqnarray}
\label{FE}
\varepsilon (\{n_{i}\}, m)=\sum_{i} f (n_{i})+m\hbar \omega \, .
\end{eqnarray}
Furthermore, demanding the periodicity condition (\ref{P}) for the
Floquet functions (\ref{SOL}) requires that the quantum number $m$
must be an integer. Thus, the quasienergy spectrum (\ref{FE})
repeats itself periodically on the energy axis.
\subsection{Time-Independent Hamiltonian}
Now we further investigate the full Floquet Hamiltonian
\begin{equation}
\hat{\cal{H}}(t) =\hat{H}(t) -i\hbar \frac{\partial}{\partial t}
\end{equation}
within the extended Hilbert space by determining its corresponding matrix elements
with respect to the Floquet functions:
\begin{eqnarray}
\hat{\cal {H}}_{\{ n'_{i}\},m';\{ n_{i}\},m}&=&
\langle\langle \{n'_{i}\} ,m' |\hat{\cal{H}}| \{n_{i}\}, m \rangle \rangle \, .
\end{eqnarray}
Using (\ref{HH}) and (\ref{SC})--(\ref{FE}) these matrix elements turn out to be
\begin{eqnarray}
&&\hat{\cal {H}}_{\{ n'_{i}\},m';\{ n_{i}\},m}=
\delta_{m,m^{'}} \left[ \sum_{i} f (n_{i})+m\hbar \omega \right] \delta_{\{ n'_{i}\},\{ n_{i}\}}
\nonumber \\ 
&&-\sum_{ij} t_{ij} \langle \{n'_{i}\}|\hat{a}^{\dag}_{i}
J_{m-m^{'}}\left( G(\hat{n}_{i},\hat{n}_{j}) \right)\hat{a}_{j}
|\{ n_{j} \} \rangle \,,
\end{eqnarray}
where $J_{m-m^{'}}$ represents a Bessel function of first kind with the argument
\begin{eqnarray}
\label{G}
G(\hat{n}_{i},\hat{n}_{j}) = \frac{g(\hat{n}_{j})-g(\hat{n}_{j}-1)+g(\hat{n}_{i})-g(\hat{n}_{i}+1)}{\hbar\omega} \, .
\end{eqnarray}
Now it is in order to take into account the physical constraints upon the driving frequency $\omega$.
On the one hand the excitation energy  $\hbar \omega$ must be much smaller than the gap between
the lowest and the first excited energy band, otherwise a single-band Bose-Hubbard model would no
longer be a valid description. On the other hand the
excitation energy $\hbar \omega$ must also be much larger than the
system parameters $t_{ij}$ and $U$, so that transitions between states with
$m \neq m'$ are highly suppressed \cite{martin2}. Thus, in the latter case
only terms with $m=m'$ have to be taken
into account, so the original time-dependent Hamiltonian (\ref{HH}) is mapped
approximately to the effective time-independent Hamiltonian 
\begin{eqnarray}
\hat{H}_{\rm eff}&=&\sum_{i}f (\hat{n}_{i}) 
-\sum_{ij}t_{ij}\hat{a}_{i}^{\dag}J_{0}\left(G(\hat{n}_{i},\hat{n}_{j})\right)\hat{a}_{j} \, .
\label{heff0}
\end{eqnarray}
Specializing (\ref{G}) and (\ref{heff0}) for the case of a periodic modulation of the s-wave scattering length (\ref{f}) and (\ref{g}) then finally yields \cite{luis}:
\begin{eqnarray}
\hat{H}_{\rm eff}&=&\sum_{i}\left[ \frac{U}{2}\hat{n}_{i}(\hat{n}_{i}-1) - \mu
\hat{n}_{i} \right]\nonumber \\
&& -\sum_{ij}t_{ij}\hat{a}_{i}^{\dag}J_{0}\left(\frac{A}{\hbar
\omega}(\hat{n}_{j}-\hat{n}_{i})\right)\hat{a}_{j} \, .
\label{heff}
\end{eqnarray}
For typical experimental parameters this means that the driving frequency $\omega$ must be of the order of kHz \cite{martin1}.
According to (\ref{heff})
we can then conclude that the effect of the time-periodic modulation of the s-wave scattering
length essentially leads to a renormalization of the hopping matrix elements. But in contrast to the shaken
optical lattice treated in Ref.~\cite{martin1}, this renormalization yields an effective conditional hopping
in the sense that it depends via the Bessel function $J_0$ on the respective particle
numbers $n_i$ and $n_j$  at the involved sites $i$ and $j$.
\section{Quantum Phase Diagram}\label{section3}
Now we determine the quantum phase diagram for the effective time-independent Hamiltonian (\ref{heff}) 
with conditional nearest neighbor hopping
at zero temperature. As the corresponding one-dimensional lattice problem has already been
discussed in detail in Ref.~\cite{luis}, we restrict us here to the higher dimensional cases of a 2D square and a 3D cubic lattice.
In order to obtain reliable results we combine different analytical approaches with
quantum Monte Carlo (QMC) simulations.
\subsection{Effective Potential Landau Theory}
In order to deal analytically with the spontaneous symmetry breaking of the inherent U(1) symmetry of bosons in an optical lattice
the Effective Potential Landau Theory (EPLT) has turned out to be quite successful
\cite{pelster1,pelster2,pelster3,tao,ohliger,mobarak}. 
Whereas the lowest order of EPLT leads
to similar results as mean-field theory \cite{Fisher:PRB89}, higher hopping
orders have recently been evaluated via the process-chain
approach \cite{martin3}, which determines the location of the
quantum phase transition  to a similar precision as demanding quantum Monte
Carlo simulations \cite{Svistunov}.
Here we follow Refs.~\cite{pelster1,tao}
and couple at first the annihilation and
creation operators to external source fields with uniform strength $j^*$ and $j$:
\begin{equation}
\hat{H}_{\rm eff}(j^{*},j)=\hat{H}_{\rm eff}+\sum_{i}\left(j^{*}\hat{a}_{i}+j\hat{a}_{i}^\dagger\right) \, .
\end{equation}
Then we calculate the ground-state energy, which coincides with the grand-canonical free energy at zero
temperature. For vanishing source fields and hopping
the unperturbed ground-state energy reads
$F_0=N_{\rm s} f(n)$ with the total number of sites $N_{\rm s}$  and the abbreviation $f(n)=U n(n-1)/2-\mu n$. In order to have $n$ particles
per site, the chemical potential has to fulfill the condition
$(n-1)<\mu/U<n$.
By applying Rayleigh-Schr\"odinger theory, we can then determine the
grand-canonical free energy perturbatively. An expansion with respect to the source fields yields
\begin{equation}
F(j,j^{*};t)=F_{0}+N_{s}\left(\sum_{p=1}^{\infty} c_{2p}(t)\left|j
\right|^{2p}\right)\,,
\label{ex}
\end{equation}
where the coefficients $c_{2p}(t)$ can be written in a power series of the hopping matrix element $t$
\begin{equation}
c_{2p}(t)=\sum_{n=0}^{\infty}(-t)^{n}\alpha_{2p}^{(n)} \, .
\label{coeff}
\end{equation}
Performing a truncation at first hopping order, we get for $p=1$
\begin{equation}
\alpha_{2}^{(0)}=\frac{n+1}{f(n)-f(n+1)}+\frac{n}{f(n)-f(n-1)}
\label{alpha0}
\end{equation}
and
\begin{eqnarray}
\alpha_{2}^{(1)}&=&z\left[\frac{n+1}{f(n)-f(n+1)}+\frac{n}{f(n)-f(n-1)}\right]^{2}
\nonumber \\
& &+z\frac{2(n+1)n\left[J_{0}\left(\frac{A}{\hbar
\omega}\right)-1\right]}{\left[f(n)-f(n+1)\right]\left[f(n)-f(n-1)\right]}\,,
\label{alpha1}
\end{eqnarray}
where $z=2D$ denotes the coordination number, i.e.~the number of
nearest neighbor sites. Note that the first line on the right-hand
side of Eq.~(\ref{alpha1}) equals to $z
\left[\alpha_{2}^{(0)}\right]^2$ due to a factorization rule for
the corresponding diagrammatic representation, see Appendix
\ref{factorization}. In contrast to this, the second line in
Eq.~(\ref{alpha1}) reveals the break down of this factorization
rule for non-vanishing driving.

As the grand-canonical free energy allows to calculate the order parameter via
\begin{eqnarray}
\psi = \frac{1}{N_{s}}\frac{\partial F}{\partial j^{*}}\,, \quad
\psi^* =  \frac{1}{N_{s}}\frac{\partial F}{\partial j}\,,
\label{psi}
\end{eqnarray}
it motivates the idea that it is possible to formally perform a
Legendre transformation from the grand-canonical free energy $F (j,j^* )$
to an effective potential $\Gamma(\psi , \psi^*)$ that is useful in a
quantitative Landau theory:
\begin{equation}
\Gamma=F/N_{\rm s}-j\psi^{*}-j^{*}\psi\,.
\end{equation}
Inserting (\ref{ex}) the effective potential
can be written in a power series of the order parameter
\begin{equation}
\Gamma=F_{0}/N_{s}-\frac{1}{c_{2}(t)}|\psi|^{2}+\frac{c_{4}(t)}{c_{2}(t)}|\psi|^{4}+\cdots \,.
\end{equation}
From (\ref{coeff}) we get the first-order result
\begin{eqnarray}
\frac{1}{c_{2}(t)}&=&\frac{1}{\alpha_2^{(0)}} \left(1+\frac{\alpha_2^{(1)}}{\alpha_2^{(0)}} t\right) \, .
\label{condL}
\end{eqnarray}
According to the Landau theory, the location of a second-order phase transition is exclusively determined
by the vanishing of (\ref{condL}), which yields with (\ref{alpha0}) and  (\ref{alpha1}):
\begin{eqnarray}
1+ zt \left[ \frac{n}{f(n)-f(n-1)} + \frac{n+1}{f(n)-f(n+1)} \right]\nonumber \\
+\frac{n(n+1)2 z t \left[J_0 \left( \frac{A}{\hbar \omega}\right)-1\right]}{[f(n)-f(n-1)][f(n)-f(n+1)]}
 =0 \, .
\label{epltresult}
\end{eqnarray}
Note that, in the special case of a vanishing driving, i.e.~$A=0$,
the quantum phase boundary (\ref{epltresult}) reduces to the
undriven mean-field result from Ref.~\cite{Fisher:PRB89} due to
$J_0(0)=1$. In order to estimate the validity of the first-order
EPLT result (\ref{epltresult}), we will compare it in the next
section with the result from the Gutzwiller mean-field theory.
\subsection{Gutzwiller Mean-Field Theory}
\begin{figure}
\includegraphics[width=0.5\textwidth]{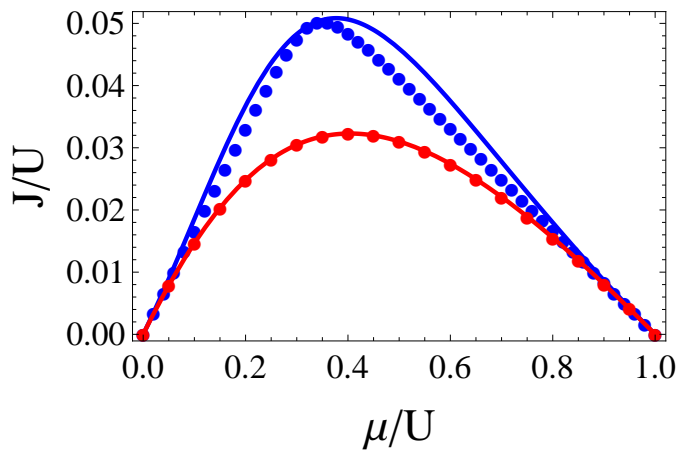}
\caption{(Color online) First-order EPLT results (\ref{epltresult}) (solid) compared with GMFT results (\ref{gmftresult}) (dots) for 3D
cubic lattice when the driving parameter
$A / (\hbar\omega)$ equals to $1$ (red) and $2.2$ (blue), respectively.}\label{figure1}.
\end{figure}

In this subsection we follow the standard Gutzwiller Mean-Field Theory (GMFT) from Refs.~\cite{gmf1,gmf2,gmf3,gmf4,gmf5,gmf6}
and assume that the ground state of the system is written as a product of identical single-site wave functions
in the basis of local Fock states
\begin{eqnarray}
| \psi \rangle = \prod_{i} \sum_{n_i=0}^\infty g_{n_i}|n_i \rangle_i \, .
\end{eqnarray}
By restricting each sum to the three states $|n_i -1 \rangle_i$, $|n_i \rangle_i$ and
$|n_i+1\rangle_i$ the ground-state energy per site results in
\begin{eqnarray}
E/N_{\rm s} &=&-zt\Bigg[\left(n+1\right) g_{n}^{2}g_{n+1}^{2}+2J_{0}\left(\frac{A}{\hbar
\omega}\right)\sqrt{n\left(n+1\right)}  \nonumber\\
& & \times g_{n}^{2} g_{n+1} g_{n+1}+n
g_{n-1}^{2}g_{n}^{2}\Bigg]+f(n-1) g_{n-1}^{2}\nonumber \\
&&+f(n)g_{n}^{2}+f(n+1) g_{n+1}^{2} \, .
\label{ground}
\end{eqnarray}
The yet unknown parameters $g_{n-1},g_n,g_{n+1}$ are then determined from
minimizing the ground-state energy (\ref{ground}) by taking into account the normalization condition
\begin{eqnarray}
g_{n-1}^2+g_n^2+g_{n+1}^2 = 1 \, .
\end{eqnarray}
Thus, the quantum phase boundary follows from the ansatz $g_{n-1}= \delta g_{n-1}$,
$g_n=1+\delta g_n$, $g_{n+1}=\delta g_{n+1}$ with infinitesimal but non-vanishing deviations $\delta g_{n-1}$, $\delta g_{n}$, $\delta g_{n+1}$,
yielding the condition
\begin{eqnarray}
1+ zt \left[ \frac{n}{f(n)-f(n-1)} + \frac{n+1}{f(n)-f(n+1)} \right]\nonumber \\
+\frac{n(n+1)z^2 t^2\left[ 1-J_0^2 \left( \frac{A}{\hbar \omega}\right)\right]}{[f(n)-f(n-1)][f(n)-f(n+1)]}
 =0 \, .
\label{gmftresult}
\end{eqnarray}
Similar to (\ref{epltresult}) also here the quantum phase boundary
(\ref{gmftresult}) yields in the limit $A \rightarrow 0$ the
undriven mean-field result from Ref.~\cite{Fisher:PRB89}. For a
non-vanishing driving, however, we observe that the quantum phase
boundary following from first-order EPLT in (\ref{epltresult}) and
GMLT in (\ref{gmftresult}) differ. In Fig.~\ref{figure1} we compare them for different driving amplitudes $A$.
We assume that the driving amplitude $A$ is restricted according to $0 < A/(\hbar\omega) < x_1$, where $x_1\approx 2.4$ denotes the first zero of the Bessel function $J_0(x)$,
so that pair superfluidity does not occur \cite{luis}.
From the comparison in Fig.~\ref{figure1} we read off that both results (\ref{epltresult}) and
(\ref{gmftresult}) are almost identical provided that the driving
parameter $A/(\hbar\omega)$ is small enough. In fact, their phase boundaries reveal only small
discrepancies for $A/(\hbar\omega) < 1.5$, whereas
qualitative different results occur for $A/(\hbar\omega) > 1.5$.
In the latter case GMFT yields a triangular lobe, while first-order EPLT predicts that the
lobe shape remains round. In comparison with GMFT and the more accurate results from the next subsection we conclude that first-order EPLT reveals for larger driving
an unphysical result insofar, as the quantum phase boundary turns out to be convex instead of concave
for small hopping $t$. Thus, the validity range of first-order EPLT is restricted up to the turning point when the convexity starts to appear. With this we obtain
from (\ref{epltresult}) irrespective of the lobe number $n$
the condition that the driving amplitude $A$ is restricted according to $0 < A/(\hbar\omega) < x_2$, where $x_2\approx1.52$ represents the smallest solution
of $2 J_0(x)=1$. In the next subsection we show that this restriction for the validity range of first-order EPLT is 
lifted once the hopping order is increased.
\begin{figure}
\includegraphics[width=0.5\textwidth]{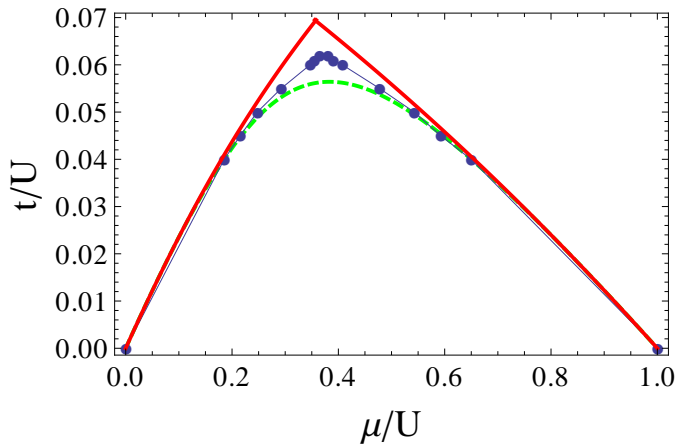}
\caption{(Color online) Phase boundary of a 2D square lattice for $J_{0}\left(A /
(\hbar \omega)\right)=0.4$: third-order strong-coupling expansion
result according to (\ref{strong1}), (\ref{strong2}) (red solid line), QMC result in the thermodynamic
limit (blue line dot), and second-order EPLT result (\ref{ep2}) (green dashed
line).} \label{figure2}.
\end{figure}
\subsection{Higher Order and Numerical Results}
Now we strive after obtaining more accurate results for the quantum phase boundary. At first, we consider EPLT in second hopping order, where
the condition for the MI-SF phase transition reads
\begin{equation}
t =\frac{\tilde{\alpha}_{1}}{2\left(\tilde{\alpha}_{2}-\tilde{\alpha}_{1}^{2}\right)}+\frac{1}{2\left(\tilde{\alpha}_{2}-\tilde{\alpha}_{1}^{2}\right)}
\sqrt{\tilde{\alpha}_{1}^{2}-4\left(\tilde{\alpha}_{2}-\tilde{\alpha}_{1}^{2}\right)}\,.
\label{ep2}
\end{equation}
Here the abbreviations
$\tilde{\alpha}_{1}=\alpha_{2}^{(1)}/\alpha_{2}^{(0)}$ and
$\tilde{\alpha}_{2}=\alpha_{2}^{(2)}/\alpha_{2}^{(0)}$ follow from
(\ref{alpha0}), (\ref{alpha1}) as well as
(\ref{alpha2})--(\ref{alpha22}) in Appendix B. In order to check
how accurate the second-order EPLT result is, we compare it
quantitatively for a 2D square lattice  with two other approaches.

On the one hand we have applied
the strong-coupling method of Ref.~\cite{monien} up to third
order, yielding for $n=1$ a quantum phase boundary with the upper
part
\begin{eqnarray}
\mu_1&=& 1 - 2z t - t^2 \left\{ 2 z^2 J_0^2  (x)
+ z \left[ \frac{3}{2}J_0^2 (2x) -6 J_0^2 (x)  \right] \right\} \nonumber\\
&& - t^3 \left\{ 6 z^3 J_0^3 (x) + z^2 \left[6 J_0 (2x) J_0^2 (x)
- 24 J_0^2 (x)\phantom{\frac{3}{2}} \right. \right. \nonumber \\
&&\left. \left. -  \frac{3}{2} J_0^2 (2x) \right] +
z \left[18 - 6 J_0 (2x) \phantom{\frac{3}{2}} \hspace*{-2mm} \right] \right\}
\label{strong1}
\end{eqnarray}
and the lower part
\begin{eqnarray}
\mu_2 &=& t z + t^2 J_0^2  (x)  (2 z^2 - 6 z) \nonumber \\
&&+ t^3 J_0^2  (x) (6 z^3 - 18 z^2 + 12 z)\, ,
\label{strong2}
\end{eqnarray}
where we have introduced for brevity the dimensionless driving parameter $x=A/(\hbar\omega)$
and the coordination number $z=4$. Note that the second-order strong-coupling results (\ref{strong1}) and (\ref{strong2})
can also be recovered from second-order EPLT by solving the quantum phase boundary $t=t(\mu)$ in (\ref{ep2})
for $\mu=\mu(t)$ up to
second hopping order. This agreement order by order in the hopping expansion
is insofar surprising as the strong-coupling method yields reliable results for low dimensions, whereas EPLT has shown
to be most accurate for high dimensions \cite{pelster1}.

In addition we have obtained high-precision QMC results from
developing an algorithm on the basis of a stochastic series
expansion \cite{sse1,sse2,qmc1,qmc2,qmc3}. 
In order to get the high accuracy quantum phase diagram in the
thermodynamic limit from QMC, we performed a finite-size scaling
with the lattice sizes $N=8\times8$, $10\times10$, and $12\times12$ at
the temperature $T=U/(20N)$. From Fig.~\ref{figure2},
we observe that the second-order EPLT result deviates not more
than $6\%$ error from the QMC result in the considered range $0 <
A/(\hbar\omega) < x_1 \approx 2.4$ of driving amplitudes. Thus,
our second-order EPLT result is sufficiently accurate for studying quantitatively
the effect of the periodic driving upon the quantum phase
transition. Furthermore, we also read off from Fig.~\ref{figure2}
that the QMC result lies between the third-order strong-coupling
result and the second-order EPLT result. This suggests to evaluate
both analytical methods to even higher hopping orders, for
instance, by applying the process-chain approach \cite{martin3}.
The true quantum phase boundary should then lie between the upper
boundary provided by the strong-coupling method and the lower
boundary from the EPLT method. This hypothesis cannot directly be
tested for a 3D lattice system as it is quite hard to get a
satisfying quantum phase diagram from QMC. However, as EPLT is
more accurate for higher-dimensional systems \cite{pelster1}, it
is suggestive that the error will even be smaller for a 3D cubic
lattice.

\begin{figure}
\includegraphics[width=0.5\textwidth]{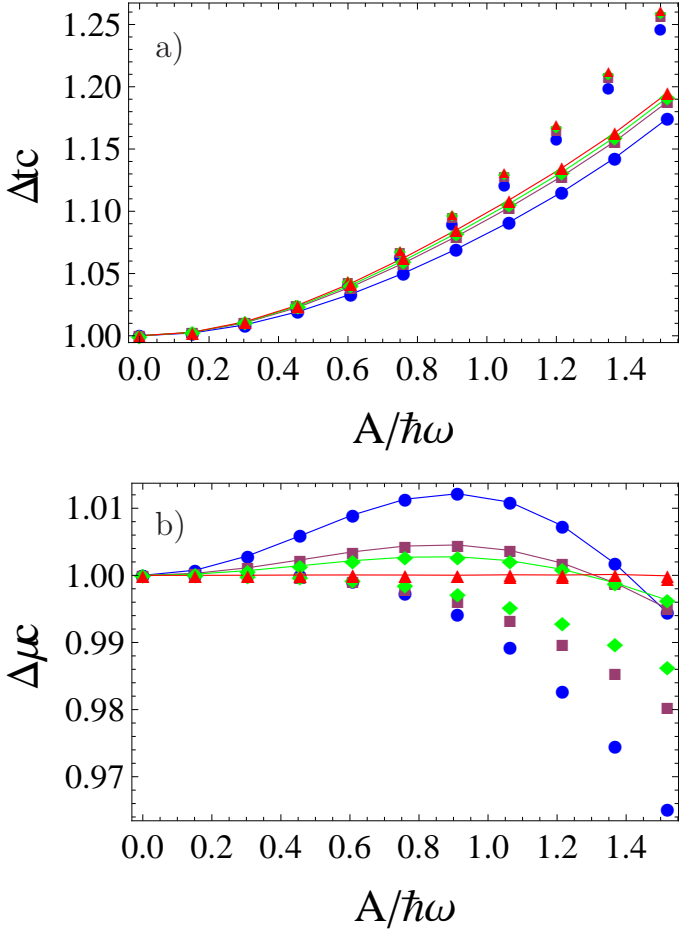}
\caption{(Color online) Relative critical a) hopping $\Delta t_c=t_{c}(A)/t_{c}(A=0)$ and
b) chemical potential $\Delta \mu_c=\left[\mu_{c}(A)-(n-1)U\right]/$ $\left[\mu_{c}(A=0)-(n-1)U\right]$
from second-order EPLT as a function of driving parameter $A/(\hbar \omega)$
for a 3D cubic lattice at tip of Mott lobes with $n=1$ (blue dot), $n=2$ (brown square dot),
$n=3$ (green diamond dot), and $n=100$ (red triangular dot) as well as for a 2D square lattice
(dots with lines).}\label{figure3}
\end{figure}

Now we use our second-order EPLT result in order to analyze
the critical points of the Mott lobes in more detail. Figure \ref{figure3} shows the EPLT predictions for the relative critical hopping
$\Delta t_c=t_{c}(A)/t_{c}(A=0)$
and the relative critical chemical potential $\Delta \mu_c=\left[\mu_{c}(A)-(n-1)U\right]/$ $\left[\mu_{c}(A=0)-(n-1)U\right]$
at the tip of different Mott lobes as a function of the driving amplitude $A$.
From Fig.~\ref{figure3} a) we read off at first
that a larger driving leads to an increase of the Mott lobe. Thus, similar to the shaken optical lattice \cite{martin1}, the periodic modulation
of the s-wave scattering length provides a control knob to tune the quantum phase transition from Mott insulator to superfluid.
Furthermore, it turns out that
the effect of periodic driving upon the critical hopping at the lobe tip is slightly
larger in $3$D than in $2$D systems,
i.e.~the driving effect is sensitive to the coordination number $z$. This can be intuitively understood from the hopping term
in the effective Hamiltonian (\ref{heff}), which contains the Bessel function $J_0$
with the nearest neighbor particle difference as its argument. Generally speaking,
higher-order hopping processes have a larger driving effect
as they have a larger probability to involve a larger nearest neighbor
particle difference, and there are more possible higher order
hopping processes in higher dimensional systems. In addition, we find from Fig.~\ref{figure3} a)
that the driving effect is quite small  with respect to the
filling number $n$ as all Mott lobes in both $2$D and $3$D lattices increase almost in the same way
for a fixed driving amplitude. Also this observation is explained by the fact that the effective Hamiltonian (\ref{heff})
depends on the nearest neighbor particle number difference rather
than on the respective particle number on each site.

We can also analyze the driving effect upon the critical chemical
potential, which is depicted in Fig.~\ref{figure3} b). At first
glance we observe that the critical chemical potential behaves
differently in a $2$D square and a $3$D cubic lattice. It
decreases monotonously in $3$D with increasing driving amplitude,
but in 2D it reveals a nonmonotonous behavior and increases
initially before it also finally decreases. Furthermore, we read
off that the critical chemical potential changes only slightly
with the driving amplitude $A$ and the filling number $n$. Up to
the driving $A/(\hbar\omega)=1$ the critical potential changes
less than one per cent, whereas for a huge filling number $n=100$
it almost does not change at all.
\begin{figure}
\includegraphics[width=0.5\textwidth]{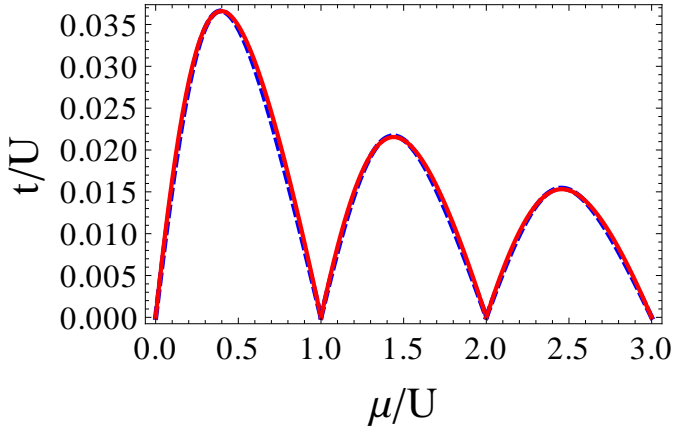}
\caption{(Color online) Quantum phase diagram of the effective model (\ref{heff}) (red solid line) and the new effective
model (\ref{neff}) (dashed blue line) with the driven parameter  $A/(\hbar \omega)=1$ for
a $3$D cubic lattice.}\label{figure4}
\end{figure}
\section{Effective Bose-Hubbard Model}\label{section4}
In the previous section we have found that the critical hopping is uniformly renormalized
according to Fig.~\ref{figure3} a) irrespective of the Mott lobe number $n$, whereas
the critical chemical nearly does 
not change according to Fig.~\ref{figure3} b). This motivates to investigate in this section whether the whole quantum phase boundary
for the effective Hamiltonian (\ref{heff}) stems approximately from
the usual Bose-Hubbard Hamiltonian
\begin{equation}
\hat{H}=-t\lambda(x)\sum_{<ij>}\hat{a}_{i}^{\dag}\hat{a}_{j}+\sum_{i}\frac{U}{2}\hat{n}_{i}(\hat{n}_{i}-1)
-\sum_{i}\mu \hat{n}_{i}\,.
\label{neff}
\end{equation}
Here $\lambda (x)$ denotes a suitable rescaling of the hopping
with the dimensionless driving parameter $x=A/(\hbar \omega)$ such that all Mott lobes coincide approximately.
From Fig.~\ref{figure3} a), we determine via a Taylor expansion for $n=1$ the
fit function
\begin{equation}
\lambda (x)=1+ax+bx^{2}+cx^{3}+dx^{4}+\ldots \,,
\label{fit}
\end{equation}
with $a=-0.0045$, $b=0.1356$, $c=0.0366$, $d=0.0129$ for a $3$D cubic lattice,
while we have $a=-0.0018$, $b=0.1212$, $c=0.0561$, $d=0.0178$ for a $2$D square lattice.
Figure \ref{figure4} compares the resulting quantum phase diagram for the new effective model (\ref{neff}) with the original effective model
(\ref{heff}). We read off that not only the critical
point but also the complete quantum phase diagram is perfectly reproduced by the Bose-Hubbard model (\ref{neff}) with the fit function (\ref{fit})
in a $3$D cubic lattice. The same can also be observed in a $2$D square lattice provided the driving amplitude is not too large.

\begin{figure}
\includegraphics[width=0.5\textwidth]{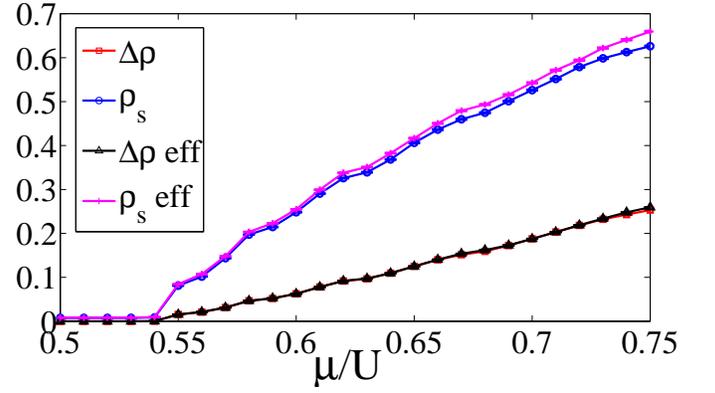}
\caption{(Color online) Total density difference and superfluid density of the
original model (\ref{heff}) and the effective model (\ref{neff})
for $t/U=0.05$, $A / (\hbar \omega)=0.4$, $N=8\times8$, and
$T=U/(20N)$ in $2$D square lattice.} \label{figure5}
\end{figure}

Furthermore, we have also used QMC simulations in order to
investigate whether both models (\ref{heff}) and (\ref{neff}) also
have the same properties in the superfluid phase. To this end we
have calculated both the superfluid density $\rho_s=\langle W^2/2\beta
t\rangle$ in terms of the winding number $W$ following Ref.~\cite{winding} and the difference between the total density and the
density in Mott-I, i.e.~$\Delta\rho=\mathop{\sum}_i \langle
n_i\rangle/N-1$, as a function of the chemical potential. 
Figure \ref{figure5} compares both quantities for the models (\ref{heff}) and (\ref{neff}).
We read off that they perfectly agree near the
quantum phase boundary, but farther away they slightly differ due to larger
density fluctuations.
\section{conclusion}
We have applied the Floquet theory in order to analyze the effect of a periodic modulation of the s-wave scattering
length upon the quantum phase diagram of bosons in an optical lattice. At first we have obtained 
a time-independent effective
Hamiltonian for large enough driving frequencies. Then we used GMFT, EPLT, strong-coupling method, and QMC simulations in order to determine quantitatively how
the different Mott lobes change with the driving amplitude. In particular, we have found that the time-independent effective model
can be well described even by the usual Bose-Hubbard model provided that the hopping is rescaled appropriately with the driving amplitude.
Thus, a period driving of the {\it interaction} 
allows to tune dynamically the {\it hopping} within an optical lattice.
Furthermore, these findings corroborate the hypothesis that the Bose-Hubbard model with a periodically driven s-wave scattering length
and the usual Bose-Hubbard model belong to the same universality class from the
point of view of critical phenomena \cite{zinn-justin,kleinert} and, thus, should have the same critical 
exponents~\cite{Fisher:PRB89,sachdev,crit1}.
\begin{acknowledgments}
We are grateful to Martin Holthaus and Marco Roncaglia for useful discussions.
Furthermore, T. Wang thanks for the financial support from the
Chinese Scholarship Council (CSC).  This work is also supported by the
German Research Foundation (DFG) via the Collaborative Research Center
SFB/TR49. Finally, both A. Pelster and T. Wang thank the Hanse-Wissenschaftskolleg for hospitality.
\end{acknowledgments}
\begin{appendix}
\section{Break Down of Factorization Rule} \label{factorization}
The perturbative coefficients $\alpha_{2p}^{(n)}$ in Eq.~(\ref{coeff})
follow from applying Rayleigh-Schr\"{o}dinger perturbation
theory by using a suitable diagrammatic representation
\cite{pelster1,tao}. By denoting the creation (annihilation) operator
with an arrow line pointing into (out of) the site, each
perturbative contribution of $\alpha_{2p}^{(n)}$ can be sketched as
an arrow-line diagram, which is composed of $n$ oriented internal
lines connecting the vertices and two external arrow lines. The
vertices in the diagram correspond to the respective lattice
sites, oriented internal lines stand for the hopping process
between sites, and the two external arrow lines are representing
creation and annihilation operators, respectively. To make this
clearer, let us consider the simplest example of the coefficient $\alpha_{2}^{(1)}$, which
has the following diagrammatic representation:
\begin{equation}
\alpha_{2}^{(1)}=
\raisebox{-0.05cm}{\includegraphics[width=3cm]{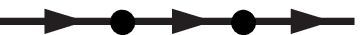}}\,.
\label{example}
\end{equation}
In the usual Bose-Hubbard model, which we recover from the Hamiltonian (\ref{heff}) for vanishing driving, i.e.~$A=0$, such a one-particle reducible diagram
reduces into its one-particle irreducible contributions in formal analogy to the Feynman diagrams of quantum field theory \cite{zinn-justin,kleinert}.
Thus, the diagrammatic representation (\ref{example}) factorizes as follows:
\begin{equation}
\alpha_{2}^{(1)} =
\raisebox{-0.03cm}{\hspace{0.05cm}\includegraphics[width=1cm]{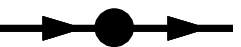}}\sum_{\langle
i,j\rangle}t_{ij}\raisebox{-0.03cm}{\hspace{0.05cm}\includegraphics[width=1cm]{figure7.eps}} \, .
\label{a2}
\end{equation}
In the latter equation '\raisebox{-0.03cm}{\hspace{0.05cm}\includegraphics[width=1cm]{figure7.eps}}'
and '$\sum_{\langle i,j\rangle}t_{ij}$' turn out to be independent, so it can be rewritten for nearest neighbor hopping according to
\begin{equation}
\alpha_{2}^{(1)} = zt
\left(\raisebox{-0.03cm}{\hspace{0.05cm}\includegraphics[width=1cm]{figure7.eps}}\right)^{2} \, .
\end{equation}
For our effective Hamiltonian (\ref{heff}), however, the coefficient $\alpha_{2}^{(1)}$ results in
\begin{equation}
\alpha_{2}^{(1)} =
\raisebox{-0.03cm}{\hspace{0.05cm}\includegraphics[width=1cm]{figure7.eps}}\sum_{\langle
i,j\rangle}t_{ij}J_{0}\left(\frac{A}{\hbar
\omega}\left(\hat{n}_{i}-\hat{n}_{j}\right)\right)\raisebox{-0.03cm}{\hspace{0.05cm}\includegraphics[width=1cm]{figure7.eps}}\,,
\label{a3}
\end{equation}
where the conditional hopping '$\sum_{\langle
i,j\rangle}t_{ij}J_{0}\left(\frac{A}{\hbar
\omega}\left(\hat{n}_{i}-\hat{n}_{j}\right)\right)$' depends on the
occupation numbers of the neighboring sites, thus it is related to the diagrams
'\raisebox{-0.03cm}{\hspace{0.05cm}\includegraphics[width=1cm]{figure7.eps}}' which appear in front and thereafter.
As a result, all three diagrammatic parts in equation (\ref{a3}) represent together one entity rather
than three independent ones, thus yielding a break down of the factorization
rule. This has the immediate consequence that one-particle reducible diagrams for non-vanishing driving will not vanish in the effective potential in any hopping order.
\section{Second Hopping Order}
Due to the break down of the factorization rule the calculation of coefficients in Eq.~(\ref{coeff}) in higher hopping orders
turns out to be much more elaborate.
For instance, the dependence of the second-order coefficient $\alpha_{2}^{(2)}$ on the coordination number $z$ 
decomposes according to
\begin{equation}
\alpha_{2}^{(2)}=z(z-1)\alpha_{21}^{(2)}+z \alpha_{22}^{(2)}\,.
\label{alpha2}
\end{equation}
The first contribution reads
\begin{eqnarray}
\alpha_{21}^{(2)}&=&\left[\frac{n}{f\left(n\right)-f\left(n-1\right)}+\frac{n+1}{f\left(n\right)-f\left(n+1\right)}\right]^{3}
\nonumber \\
& &+\left[J_{0}\left(\frac{A}{\hbar
\omega}\right)+3\right]\left[J_{0}\left(\frac{A}{\hbar
\omega}\right)-1\right] \label{alpha21} \\
& &\times \left\{\frac{n\left(n+1\right)^{2}}{\left[f(n)-f(n-1)\right]\left[f(n)-f(n+1)\right]^{2}}
\right. \nonumber \\
& &\left.
+ \frac{n^{2}\left(n+1\right)}{\left[f(n)-f(n-1)\right]^{2}\left[f(n)-f(n+1)\right]}\right\}
\nonumber
\end{eqnarray}
whereas the second term turns out to be
\begin{widetext}
\begin{eqnarray}
\alpha_{22}^{(2)}&=&
\frac{n^{3}}{\left[f(n)-f(n-1)\right]^{3}}
+\frac{\left(n+1\right)^{3}}{\left[f(n)-f(n+1)\right]^{3}}
-\frac{(n-1)n(n+1)J_{0}^{2}\left(\frac{2A}{\hbar \omega}\right)}{\left[f(n)-f(n-1)\right]^{2}\left[f(n+1)+f(n-2)-2f(n)\right]}
\nonumber \\&&
+\frac{(n-1)n(n+1) J_{0}^{2}\left(\frac{A}{\hbar \omega}\right)}{\left[f(n-1)-f(n)\right]\left[f(n-1)+f(n+1)-2f(n)\right]^{2}}
-\frac{n(n+1)(n+2) J_{0}^{2}\left(\frac{A}{\hbar \omega}\right)}{\left[f(n)-f(n+1)\right]\left[f(n-1)+f(n+1)-2f(n)\right]^{2}}\nonumber \\
&&
-\frac{(n-1)n(n+1)J_{0}^{2}\left(\frac{A}{\hbar \omega}\right)}{\left[f(n+1)+f(n-2)-2f(n)\right]\left[f(n-1)+f(n+1)-2f(n)\right]^{2}} \nonumber\\
&&+\frac{2n^{2}(n+1)\left[J_{0}^{2}\left(\frac{A}{\hbar \omega}\right) 
-J_{0}\left(\frac{A}{\hbar \omega}\right)\right]}{\left[f(n-1)-f(n)\right]^{2}\left[f(n-1)+f(n+1)-2f(n)\right]}
\nonumber \\
&&+\frac{2n(n+1)^{2}\left[J_{0}^{2}\left(\frac{A}{\hbar \omega}\right)-J_{0}\left(\frac{A}{\hbar
\omega}\right)\right]}{\left[f(n)-f(n+1)\right]^{2}\left[f(n-1)+f(n+1)-2f(n)\right]}
-\frac{n(n+1)(n+2)J_{0}^{2}\left(\frac{2A}{\hbar \omega}\right)}{\left[f(n)-f(n+1)\right]^{2}\left[f(n-1)+f(n+2)-2f(n)\right]}
\nonumber \\
&&-\frac{2(n-1)n(n+1)J_{0}\left(\frac{A}{\hbar \omega}\right)J_{0}\left(\frac{2A}{\hbar
\omega}\right)}{\left[f(n-1)-f(n)\right]\left[f(n-2)+f(n+1)-2f(n)\right]\left[f(n-1)+f(n+1)-2f(n)\right]}
\nonumber \\
&&-\frac{n(n+1)(n+2)J_{0}^{2}\left(\frac{A}{\hbar \omega}\right)}{\left[f(n+1)+f(n-1)-2f(n)\right]^{2}\left[f(n-1)+f(n+2)-2f(n)\right]}
\nonumber \\
&&+\frac{2n(n+1)(n+2)J_{0}\left(\frac{A}{\hbar \omega}\right)J_{0}\left(\frac{2A}{\hbar
\omega}\right)}{\left[f(n)-f(n+1)\right]\left[f(n+2)+f(n-1)-2f(n)\right]\left[f(n-1)+f(n+1)-2f(n)\right]}\,.
\label{alpha22}
\end{eqnarray}
\end{widetext}
\end{appendix}
\end{document}